\documentclass[journal]{IEEEtran}
\usepackage{amsfonts}
\usepackage{amsmath}
\usepackage{amssymb}
\usepackage{lscape}
\usepackage{epsf}
\usepackage{graphicx}

\newtheorem{theorem}{\indent Theorem}[section]
\newtheorem{lemma}[theorem]{\indent Lemma}

\newtheorem{EXAMPLE}{\indent Example}[section]

\newcommand{\code}{{\mathcal{C}}}

\newcommand{\cE}{{\mathcal{E}}}
\newcommand{\decoder}{{\mathcal{D}}}

\newcommand{\cB}{{\mathcal{B}}}
\newcommand{\cA}{{\mathcal{A}}}

\newcommand{\cN}{{\mathcal{N}}}
\newcommand{\cK}{{\mathcal{K}}}
\newcommand{\cP}{{\mathcal{P}}}

\newcommand{\ho}{{h_{\mbox{\scriptsize o}}}}

\newcommand{\ff}{{\mathbb{F}}}

\newcommand\nn{{\mathbb N}}

\newcommand{\blde}{{\mbox{\boldmath $e$}}}
\newcommand{\bldalpha}{{\mbox{\boldmath $\alpha$}}}

\newcommand{\bldv}{{\mbox{\boldmath $v$}}}
\newcommand{\bldu}{{\mbox{\boldmath $u$}}}
\newcommand{\bldw}{{\mbox{\boldmath $w$}}}

\newcommand{\bldzero}{{\mbox{\boldmath $0$}}}

\newcommand{\qed}{\hspace*{\fill}%
    \vbox{\hrule\hbox{\vrule\squarebox{.667em}\vrule}\hrule}\smallskip}
    \def\squarebox#1{\hbox to #1{\hfill\vbox to #1{\vfill}}}

\newlength{\Initlabel}
\newlength{\Algwidth}

\begin{document}

\title{Recursive Code Construction for Random Networks
\author{Vitaly Skachek,~\IEEEmembership{Member,~IEEE}
\thanks{%
    This work was supported by the Claude Shannon Institute for 
    Discrete Mathematics, Coding and Cryptography (Science Foundation Ireland Grant 06/MI/006).} 
\thanks{%
    V. Skachek is with the Division of Mathematical Sciences, School of Physical and Mathematical Sciences,
    Nanyang Technological University, 21~Nanyang Link, 637371, Singapore (e-mail: Vitaly.Skachek@ntu.edu.sg). 
    This work was done while he was with the Claude Shannon Institute
    and the School of Mathematical Sciences, University College Dublin, Belfield, Dublin 4, Ireland.}      
 }
}



\maketitle

\begin{abstract}

A modification of K\"otter-Kschischang codes for random networks is presented
(these codes were also studied by Wang \emph{et al.} 
in the context of authentication problems). 
The new codes have higher information rate, while maintaining the same 
error-correcting capabilities. An efficient error-correcting algorithm is 
proposed for these codes.  

\end{abstract}


\section{Background}

The area of network coding has emerged since the work of Ahlswede {\em et al.}~\cite{Ahlswede}. 
It was shown that sending coded information over the network yields an advantage in 
bandwidth utilization compared with the classical routing scenario. Furthermore, the use of
network coding for correcting errors in the information sent over the network was suggested in~\cite{Cai}.
This approach relied, however, on the knowledge of the network topology. 

In~\cite{KK}, a new approach for error-correcting network coding was suggested. It was assumed that the 
network topology is not known. The encoded information was represented by subspaces of some 
vector space. So-called \emph{codes for random networks} were constructed and
a corresponding decoder was presented in that work, that uses some ideas from 
classical Reed-Solomon codes and decoders. Essentially, the analogous construction was proposed several years 
earlier in the context of authentication codes~\cite{Naini}. 
 
In~\cite{Gadouleau} and~\cite{SKK}, the connections between matrix codes in the rank metric~\cite{Gabidulin} 
(see also~\cite{Roth}) and the codes for random networks
were established. In particular, the code construction in~\cite{KK}
is a {\em lifting} of the matrix codes in~\cite{Gabidulin}. 

A decoder for the codes in~\cite{KK} was proposed therein, and an alternative (more efficient) 
decoder for the same codes was proposed in~\cite{SKK}. 
A construction of related family of 
\emph{spread codes} was presented in~\cite{Felice} (similar construction appeared independently also in~\cite{Bossert}). 

Recently, a generalization of codes in~\cite{KK} was presented in~\cite{Silberstein}. 
In the present work, we also modify the construction in~\cite{KK}. 
Our codes have 
the same error-correcting capability as the codes in~\cite{KK}, while the 
information rate of our codes is higher. 
Present construction can be viewed as a generalization of the construction in~\cite{Felice}.
On the other hand, this construction can be viewed as a special case of the construction 
in~\cite{Silberstein}. 
An efficient decoding algorithm is provided
for the codes presented in this work, 
it uses the decoder in~\cite{KK} or~\cite{SKK} as a subroutine.      
By contrast, no efficient decoding algorithm for the codes in~\cite{Silberstein} was given.

It should be mentioned that the present work was done 
independently of~\cite{Silberstein}, and about at the same time. 
The more updated version of~\cite{Silberstein} was published recently as~\cite{Etzion-Silberstein}.

\section{Notations and Previous Results}

Let $W$ be a vector space over a finite field $\ff_q$ and let $V, U \subseteq W$ be linear subspaces of $W$. 
We use the notation $\dim(V)$ for the dimension of $V$. We denote 
the sum of $U$ and $V$ as $U+V = \{ \bldu + \bldv \; : \; \bldu \in U, \bldv \in V \}$. If $U \cap V = \emptyset$, then
for any $\bldw \in U + V$ there is a unique representation $\bldw = \bldu + \bldv$, where $\bldu \in U$ and 
$\bldv \in V$. In this case we say that $U+V$ is a direct sum, and denote it as $U \oplus V$. 
It is easy to check that $\dim(U \oplus V) = \dim(U) + \dim(V)$.  
For a vector set $S \subseteq W$, 
we use the notation $\mbox{span} ( S )$ to denote a linear span of the vectors in $S$. 
We use the notation $\bldzero^m$ to denote all-zero vector of length $m$, for any integer $m$.
When the value of $m$ is clear from the context, we sometimes write $\bldzero$ rather than $\bldzero^m$. 

Below, we recall the construction in~\cite{KK}.
Let $\ff = \ff_{q^m}$ be an extension field of $\ff_q$, $m > 1$. Then, $\ff$ is
a vector space over $\ff_q$. Let $\{ \bldalpha_1, \bldalpha_2, \cdots, \bldalpha_\ell \} \subseteq \ff$ 
be a set of linearly independent elements in $\ff$. 
We denote 
\[
\langle A \rangle = \mbox{span} \Big( \{ \bldalpha_1, \bldalpha_2, \cdots, \bldalpha_\ell \} \Big) 
\]
and 
\[
W = \langle A \rangle \oplus \ff 
= \left\{ ( \bldv_1, \bldv_2) \; : \; \bldv_1 \in \langle A \rangle, \bldv_2 \in \ff \right\} \; . 
\]
For a vector $\bldv \in W$, sometimes we may write $\bldv = (\bldv_1, \bldv_2)$, where 
$\bldv_1 \in \langle A \rangle$ and $\bldv_2 \in \ff$
(any such vector $\bldv$ can be viewed as an $(\ell+m)$-tuple over $\ff_q$). 
For a vector space $V \subseteq W$ we define 
a \emph{projection of $V$ on $\ff$} as
\[
V |_\ff = \left\{ \bldv_2 \in \ff \; : \; (\bldv_1, \bldv_2) \in V \right\} \; .
\]
We use the notation $\cP(W,\ell)$ for the set of all subspaces of $W$ of dimension $\ell$. 
For $U, V \in W$, 
let $d(U, V) = \dim(U) + \dim(V) - 2 \dim(U \cap V)$ be a distance between $U$ and $V$ in the Grassmanian metric 
(see~\cite{KK}).

Let $\ff^k[x]$ denote the set of linearized polynomials over $\ff$ of degree at most $q^{k-1}$, $k \ge 1$. 
Define the mapping $\cE : \ff^k[x] \rightarrow \cP(W,\ell)$ as
\[
\cE(f(x)) = \mbox{span} \Big( \left\{ (\bldalpha_1, f(\bldalpha_1)), \cdots, (\bldalpha_\ell, f(\bldalpha_\ell)) \right\} \Big) \; . 
\] 
The code $\cK$ is defined in~\cite{KK} (for $k \le \ell)$ as 
\[
\cK = \left\{ \cE(f(x))  \; : \; f(x) \in \ff^k[x] \right\} \; .   
\]
Below, we might sometimes use the notation $\cK[\ell+m, \ell, k]$ for the code $\cK$ with the parameters as above.   
It was shown in~\cite{KK} that $|\cK| = q^{mk}$. 
It was also shown there, that for all $U, V \in \cK$, $U \neq V$, it holds that
$\dim(U \cap V) \le k-1$, and so $d(U,V) \ge 2 (\ell - k + 1)$. Therefore,
the minimum distance of $\cK$ is at least $2 (\ell - k + 1)$.   

Singleton-type 
upper bound on the maximum size of $\cK$, $\cA_q(\ell+m, \ell, k)$, was derived in~\cite{KK}. The bound can be written as 
\begin{equation}
\cA_q(\ell+m, \ell, k) \le \left[ 
\begin{array}{c} 
m + k \\ 
k 
\end{array} 
\right]_q \; ,
\label{eq:singleton}
\end{equation}
where
\[
\left[ \begin{array}{c} m + k \\ k \end{array} \right]_q = \prod_{i=0}^{k-1} \frac{q^{m+k-i} - 1}{q^{k-i} - 1} \; . 
\]
The following bound was presented in~\cite{Naini} (it is always tighter than its counterpart~(\ref{eq:singleton})):
\[
\cA_q(\ell+m, \ell, k) \le 
\left[ 
\begin{array}{c} 
m + \ell \\ 
k
\end{array} 
\right]_q 
\Bigg/ \left[ 
\begin{array}{c} 
\ell \\ 
k 
\end{array} 
\right]_q 
\; . 
\]
 
Finally, the Johnson-type bound was presented in~\cite{Xia}, as below
\begin{multline*}
\cA_q(\ell+m, \ell, k) \le \\
\Bigg\lfloor \frac{q^{\ell + m}-1}{q^\ell-1} 
\Bigg\lfloor \frac{q^{\ell + m - 1}-1}{q^{\ell-1}-1} \cdots 
\Bigg\lfloor \frac{q^{\ell+m-k+1}-1}{q^{\ell-k+1}-1}
\Bigg\rfloor \cdots 
\Bigg\rfloor \Bigg\rfloor \; . 
\end{multline*}

It can be seen that there is a gap between the upper bounds and the actual size of $\cK$. 
In this work, we construct codes with a larger number of words (compared with $\cK$).

Let $V \in \cK$ and $U \in W$.
In~\cite{KK},~\cite{SKK}, the decoding algorithms for the code $\cK$ were presented, 
such that if $d(U,V) < \ell-k+1$, the algorithms applied to the input $U$ will return $V$. 
The time complexity of the algorithm in~\cite{SKK} is $O(m \cdot (\ell - k) )$. 
In this work, we present a decoding algorithm for the proposed codes with the
same decoding radius. The decoding complexity of this algorithm is 
$O\left(\frac{m^2 (\ell-k)}{\ell}\right)$ operations over $\ff_{q^m}$.  

\section{Code Construction}

In this section, we define a new code, based on the construction in~\cite{KK},~\cite{Naini}. Let $m \ge \ell \ge k$. 
We use the notation $\code[\ell+m,\ell,k]$ (for a sake of simplicity, sometimes we will use the notation $\code$
instead) to define this new random network code defined by vector subspaces  
in the ambient space W of dimension $\ell+m$, such that for any $V \in \code$, $\dim(V) = \ell$, and 
for any $V \in \code$ and $U \in \code$, $\dim(U \cap V) \le k-1$. 
Let $\cN(\ell+m, \ell, k) = \big|\code[\ell+m,\ell,k]\big|$ for all $m, \ell, k$.  

We pick code parameters $m$, $\ell$ and $k$. 
Let $h_{\ell+m}$ be an integer, $0 \le h_{\ell+m} \le k-1$. 
This $h_{\ell+m}$ is a design parameter which can be optimized later. 

Next, we recursively define the code $\code[\ell+m, \ell, k]$. 
\begin{itemize}
\item{\em Boundary condition:}
$m < 2(\ell-h_{\ell+m})$ or $\ell-h_{\ell+m} < k$
(in which case $\cK[m, \ell-h_{\ell+m},k]$ is not defined). We define $\code = \cK[\ell+m,\ell,k]$. 
\item{\em Recursive step:} assume that 
\begin{multline*}
\code[m, \ell-h_{\ell+m}, k] = \\
\{ U_\sigma \subseteq \ff : \sigma = 1, 2, \cdots, \cN(m, \ell-h_{\ell+m}, k) \} \; . 
\end{multline*}
Let 
$\{\blde^\sigma_1, \blde^\sigma_2, \cdots, \blde^\sigma_{\ell-h_{\ell+m}} \} \subseteq \ff$ be a basis of~$U_\sigma$. 
\begin{itemize}
\item
If $h_{\ell+m} = 0$, then we set $S_\sigma = \emptyset$ for all $\sigma \in \nn$. 
\item
Otherwise, for $\sigma = 1, 2, \cdots, \lfloor \ell/h_{\ell+m} \rfloor$, we define sets of vectors
\begin{multline*}
S_\sigma = \Big\{(\bldalpha_j, \bldzero^m) : \\ j = (\sigma-1)h_{\ell+m}+1, (\sigma-1)h_{\ell+m}+2, \cdots, 
\sigma h_{\ell+m} \Big\} \; .  
\end{multline*}
\end{itemize}
In addition, for $\sigma = 1, 2, \cdots, \cN(m, \ell-h_{\ell+m}, k)$, we define 
\[
T_\sigma = \left\{(\bldzero^\ell, \blde^\sigma_j) : j = 1, 2, \cdots, \ell-h_{\ell+m} \right\} \; .
\]

Let 
\begin{equation}
t_{\ell+m} = 
\left\{ \begin{array}{l}
\min \left\{ \Big\lfloor \frac{\ell}{h_{\ell+m}} \Big\rfloor, 
\cN(m, \ell-h_{\ell+m}, k) \right\} \\ \hspace{22ex} \mbox{ if } h_{\ell+m} > 0 \\
\cN(m, \ell, k)  \mbox{ if } h_{\ell+m} = 0 
\end{array} \right.  
\label{eq:t}
\end{equation}

We define vector spaces $V_\sigma \in \cP(W,\ell)$, for $\sigma = 1, 2, \cdots, t_{\ell+m}$, as
$V_\sigma = \mbox{span} \left( S_\sigma \cup T_\sigma \right). $
We also define a set 
$\cB = \{ V_\sigma \}_{\sigma = 1, 2, \cdots, t_{\ell+m}}$. 
Finally, we define a code $\code$ as $\code = \cK[\ell+m, \ell,k] \cup \cB$. 
\end{itemize}

\section{Code Parameters} 

\subsection{Recursive Formula for the Number of Codewords}
The code $\code$ as above is obviously a set of subspaces of dimension $\ell$ in the space $W$ of dimension $\ell+m$. 
The number of codewords in $\code$ is given by the recursive relation
\begin{equation}
\cN(\ell+m, \ell, k) =  q^{mk} + t_{\ell+m} \; , 
\label{eq:N}
\end{equation}
where $t_{\ell+m}$ is as in~(\ref{eq:t}).
The boundary conditions are
\begin{multline}
\cN(\ell+m,\ell,k) = \Big| \cK[\ell+m, \ell, k] \Big| = q^{mk} \\
\mbox{ if } m < 2(\ell - h_{\ell+m}) 
\mbox{ or } \ell - h_{\ell+m} < k \; .   
\label{eq:N-2}
\end{multline}
Please note that if $\cK[m, \ell-h_{\ell+m}, k] \neq \emptyset$, 
we get $\cN(\ell+m,\ell,k) > q^{mk}$, thus having more codewords compared with the construction in~\cite{KK}. 
Otherwise, we have $\cN(\ell+m,\ell,k) = q^{mk}$.

\subsection{Optimization of $h_{\ell+m}$}
Next, let us discuss the choice of the parameter $h_{\ell+m}$. We are interested in 
$h_{\ell+m}$ that maximizes $t_{\ell+m}$ in~(\ref{eq:t}).  
Note, however, that  
\[
\cN(m, \ell-h_{\ell+m}, k) \ge q^{(m - \ell + h_{\ell+m})\cdot k} \; , 
\]
where the right-hand side is an increasing function of $h_{\ell+m}$. 
For fixed $m$, $\ell$ and $k$, the function $\lfloor \ell/h_{\ell+m} \rfloor$ does not increase with $h_{\ell+m}$. 
Therefore, we have the following lemma. 

\begin{lemma}
If for some $h = 0, 1, \cdots, k-2$, 
\[
\Bigg\lfloor \frac{\ell}{h+1} \Bigg\rfloor \le q^{(m - \ell + h)\cdot k} \; , 
\]
then there exists an optimal choice of $h_{\ell+m}$ (corresponding to the maximum number of codewords) 
satisfying $h_{\ell+m} \le h$. 
\label{lemma:ell}
\end{lemma}

{\em Proof.} 
The number of codewords corresponding to the selection $h_{\ell+m} \ge h + 1$ is at most 
$\lfloor \frac{\ell}{h+1} \rfloor$. The number of codewords corresponding to the selection 
$h_{\ell+m} = h$ is either $\lfloor \frac{\ell}{h} \rfloor$ or $\cN(m, \ell-h, k) \ge q^{(m - \ell + h)\cdot k}$. 
Both expressions are greater or equal to $\lfloor \frac{\ell}{h+1} \rfloor$. \qed 

The following theorem shows that in most of the cases the optimal value of $h_{\ell+m}$ is zero. 
\begin{theorem}
Let $\ell+m$, $\ell$ and $k$ be the parameters of the code $\code$, and let $h_{\ell+m} = \ho$ be the smallest value that 
maximizes the number of words in $\code$. In addition, assume that one of the following holds: 
\begin{itemize}
\item[(C1)]
$\ell \ge 4$, $k=1$;  
\item[(C2)]
$\ell \ge 3$, $k \ge 2$. 
\end{itemize}
If $\ho > 0$, then $m < 2(\ell-\ho)$, and so $\code[m, \ell-\ho,k]$ does not exist. 
\label{thrm:contradiction}
\end{theorem}

{\em Outline of the proof.} 
Consider the code $\code[\ell+m, \ell, k]$. Assume that it is constructed recursively from the code
$\code[m, \ell-\ho, k]$ for the smallest optimal $h_{\ell+m} = \ho$, $0 < \ho \le k-1$, 
as it was described in the previous section. 
The later code exists only if 
\begin{equation}
m \ge 2 (\ell-\ho) \; . 
\label{eq:m}
\end{equation} 
Then, $0, 1, \cdots, \ho-1$ 
are all not optimal values of $h_{\ell+m}$, and so for $h = 0, 1,\cdots, \ho-1$
we have 
\[
\lfloor \ell/ (h+1) \rfloor > q^{(m - \ell + h)\cdot k} \; 
\]
due to Lemma~\ref{lemma:ell}.  
This can be rewritten (by using~(\ref{eq:m})) as
\begin{equation}
\log_q (\lfloor \ell/ (h+1) \rfloor) > (\ell - 2 \ho + h)\cdot k \; . 
\label{eq:contradiction}
\end{equation} 

Next, one has to show that for any $\ho$, $0 < \ho \le \ell-1$, there exists some $h$, $0 \le h < \ho$,  
such that~(\ref{eq:contradiction}) 
does not hold. This can be shown by taking $h = \ho - 1$. We omit the details due to their technicality.  
 
This leads to a contradiction, which follows from the assumption~(\ref{eq:m}), and
so the only possible situation is $m < 2( \ell - \ho)$. 
\qed

\subsection{Explicit Formula for the Number of Codewords}

If all $h_{\ell+m}$'s in the construction are zeros  
for all $\ell$ and $m$, then $t_{\ell+m} = \cN(m, \ell, k)$. Therefore,~(\ref{eq:N}) becomes     
\[
\cN(\ell+m, \ell, k) =  q^{mk} + \cN(m, \ell, k) \; , 
\]
and thus
\begin{eqnarray*}
\cN(\ell+m, \ell, k) & = & q^{mk} + q^{(m-\ell)k} + \cdots + q^{(r + \ell) k} \\
& = & \frac{q^{(\ell+m)k} - q^{(r + \ell) k}}{q^{\ell k} - 1} \; ,
\end{eqnarray*}
where $r = m \mod \ell$.

\subsection{Minimum Distance}

The next theorem provides a lower bound on the minimum distance of the code $\code$. 
\begin{theorem}
The minimum distance of the code $\code[\ell+m, \ell,k]$ is $2(\ell-k+1)$. 
\end{theorem}

{\em Outline of the proof.} Since $\code \subseteq \cP(W, \ell)$, 
it would be enough to show that for any two $U, V \in \code$, $U \neq V$, it holds
\[
\dim(U \cap V) \le k - 1 \;  . 
\]

Recall that $\code = \cK \cup \cB$. 
There are three cases.
\begin{enumerate}
\item $U, V \in \cK$. 
In this case, the proof follows from~\cite{KK}.
\item $U \in \cK, V \in \cB$. 

Take any $\bldv \in U \cap V$, $\bldv \neq \bldzero$. Since $\bldv \in U$, we write 
for some $f(x) \in \ff^k[x]$ and $a_i \in \ff_q$ (for $i = 1, 2, \cdots, \ell$): 
\begin{eqnarray*}
\bldv & = & \sum_{i=1}^{\ell} a_i \cdot (\bldalpha_i, f(\bldalpha_i)) \nonumber  \; . 
\end{eqnarray*}
Since $\bldv \in V$, we write 
for some $\sigma$, $1 \le \sigma \le t_{\ell+m}$, 
for some $b_i \in \ff_q$ (for $i = 1, 2, \cdots, h_{\ell+m}$) and $c_i \in \ff_q$ (for $i = 1, 2, \cdots, \ell-h_{\ell+m}$):
\begin{eqnarray*}
\bldv & = & \sum_{i=1}^{h_{\ell+m}} 
b_i \cdot (\bldalpha_{(\sigma-1)h_{\ell+m}+i}, \bldzero^m) \nonumber \\
& + & \sum_{i=1}^{\ell-h_{\ell+m}} c_i \cdot (\bldzero^\ell, \blde^\sigma_i) \nonumber \; .
\end{eqnarray*}

It follows that 
\[
a_i = \left\{ \begin{array}{ll} 
b_{i-(\sigma-1)h_{\ell+m}} \\
\hspace{5ex} \mbox{ if } (\sigma - 1) h_{\ell+m} + 1 \le i \le \sigma h_{\ell+m} \\
0 \;\; \mbox{ otherwise } 
\end{array}
\right. 
\]
We obtain that every $\bldv \in U \cap V$ can be written as 
\[
\bldv = \sum_{i=1}^{h_{\ell+m}} b_i \cdot (\bldalpha_{(\sigma-1)h_{\ell+m}+i}, f(\bldalpha_{(\sigma-1)h_{\ell+m}+i})) \; . 
\]
Therefore, $\dim(U \cap V) \le h_{\ell+m} \le k-1$. 

\item $U, V \in \cB$. Similarly to Case (2),  
take any $\bldv \in U \cap V$, $\bldv \neq \bldzero$. 
Since $\bldv \in U \cap V$, for some $\sigma \neq \tau$, $1 \le \sigma, \tau \le t_{\ell+m}$,
for some $a_i, b_i \in \ff_q$ (for $i = 1, 2, \cdots, h_{\ell+m}$) and 
for some $c_i, d_i \in \ff_q$ (for $i = 1, 2, \cdots, \ell-h_{\ell+m}$):
\begin{align*} 
& \bldv \quad = \\
& \sum_{i=1}^{h_{\ell+m}} a_i \cdot (\bldalpha_{(\sigma-1)h_{\ell+m}+i}, \bldzero^m) 
+ \sum_{i=1}^{\ell-h_{\ell+m}} d_i \cdot (\bldzero^\ell, \blde^\sigma_i) = \\ 
\end{align*}
\begin{align*}
& \sum_{i=1}^{h_{\ell+m}} b_i \cdot (\bldalpha_{(\tau-1) h_{\ell+m} +i}, \bldzero^m)
+ \sum_{i=1}^{\ell-h_{\ell+m}} c_i \cdot (\bldzero^\ell, \blde^\tau_i) \; . 
\end{align*}
 
We obtain 
that all $a_i = b_i = 0$ for $i = 1, 2, \cdots, h_{\ell+m}$. 
We also obtain that $\bldv = ( \bldzero^{\ell}, \bldu )$, where $\bldu \in U_\sigma \cap U_\tau$. Since, for all $\sigma$ and $\tau$, 
$\dim(U_\sigma \cap U_\tau) \le k-1$, it follows that $\dim(U \cap V) \le k-1$. 
\end{enumerate}
\qed

\section{Decoding} 

\subsection{Simple Case}
 
Below, we present a recursive decoding algorithm for the code $\code$, when 
$h_{i} = 0$ (for all $i$). 

Decoding algorithms for the code $\cK$ were presented in~\cite{KK} and~\cite{SKK}. 
Suppose that $V \in \cK$ is transmitted over the operator channel (see~\cite{KK} for details). 
Suppose also that an $(\ell-\kappa+\gamma)$-dimensional subspace $U$ of $W$ is received, where
$\dim(U \cap V) = \ell - \kappa$. 
We use a modification of the decoding algorithm in~\cite{SKK} as follows. 
Given a received vector space $U$ of $W$,
the decoder is able to recover a single $V \in \cK$ whenever $\kappa+\gamma < \ell-k+1$. 
If the decoding fails, the decoder returns a special error message `?' (such a modification is straight-forward). 
We will denote this decoder $\decoder^\cK_{\ell+m,\ell,k} : W \rightarrow \cK[\ell+m, \ell,k] \cup \{ \mbox{`?'} \}$.

Now, suppose that $V \in \code[\ell+m,\ell,k]$ is transmitted over the operator channel, and 
an $(\ell-\kappa+\gamma)$-dimensional subspace $U$ of $W$ is received, and 
$\dim(U \cap V) = \ell - \kappa$. 
We will denote the decoder for the code $\code[\ell+m,\ell,k]$ 
as $\decoder_{\ell+m,\ell,k} : W \rightarrow \code \cup \{ \mbox{`?'} \}$. 
The decoder is summarized in Figure~\ref{fig:alg}. If the decoding fails, the decoder returns an error message `?'.  
As we show in the sequel, 
the decoder $\decoder_{\ell+m,\ell,k}$ is able to recover $V \in \code$ from $U \in W$ given that
$\kappa+\gamma < \ell-k+1$.

\begin{figure}[hbt]
\makebox[0in]{}\hrulefill\makebox[0in]{}
\begin{description}
\item[\bf Input:] $\,$ received word $U \in W$ 

\settowidth{\Initlabel}{\textbf{Input:}}
\item[\bf Let]
$Z \leftarrow \decoder^\cK_{\ell+m, \ell, k}(U)$  
\item[\bf If] $Z =$ `?' {\bf and } $m \ge 2 \ell$ {\bf then } $\{$ \\
  $\phantom{aa}$ {\bf Let} $U' \leftarrow U |_{\ff}$  \\
  $\phantom{aa}$ {\bf Let} $V' \leftarrow \decoder_{m, \ell, k} (U')$  \\
  $\phantom{aa}$ {\bf If } $V' \neq$ `?' {\bf then} $Z \leftarrow \{ \bldzero^\ell \} \oplus V'$   \\
$\}$ 
\item[\bf If] $Z =$ `?' {\bf then} return `?' \\
\hspace{1ex} {\bf otherwise} return $Z$ 
\end{description}
\makebox[0in]{}\hrulefill\makebox[0in]{}
\caption{Decoder $\decoder_{\ell+m, \ell, k}$ for the code $\code$.}
\label{fig:alg}
\end{figure}	

\begin{theorem}
Let $V$ be transmitted over the operator channel and let $U$ be received.
In addition, let $\dim(V) = \ell$, $\dim(U) = \ell - \kappa + \gamma$, $\dim(U \cap V) = \ell - \kappa$. 
Then, the decoder in Figure~\ref{fig:alg} is able to recover the original codeword $V \in \code$ from $U$ 
given that $\kappa+\gamma < \ell-k+1$. 
\end{theorem}

{\em Proof.} There are two cases. If $V \in \cK$, then the claim follows from the correctness of 
the decoder in~\cite{SKK}. Therefore, we assume that $V \in \cB$. In that case, by the definition of $\cB$, 
\[
V = \mbox{span} \Big( \{ (\bldzero, \blde^\sigma_1), (\bldzero, \blde^\sigma_2), \cdots, (\bldzero, \blde^\sigma_\ell) 
\} \Big) = \{ \bldzero^\ell \} \oplus V' \; , 
\]
where 
$V' = \mbox{span} \Big( \{ \blde^\sigma_1, \blde^\sigma_2, \cdots, \blde^\sigma_\ell \} \Big) \in \code[m, \ell,k]$
for some $\sigma$. 
In particular, $V$ is isomorphic to $V'$ and $\dim(V) = \dim(V')$. 

Let $U' = U|_\ff$. Obviously, $\dim(U') \le \dim(U) = \ell - \kappa + \gamma$. 
We also have that
$\ell - \kappa = \dim(U \cap V) \le \dim(U' \cap V')$ (in particular, for any $\bldv \in U \cap V$, the projection of
$\bldv$ on its last $m$ coordinates lies in $U' \cap V'$, and for $\bldv \neq \bldu \in U \cap V$, the 
projections of $\bldu$ and $\bldv$ on the last $m$ coordinates yield different vectors, since the only puctured 
coordinates are zero-coordinates). 

We have 
\begin{eqnarray*}
d(U', V') & = & \dim(U') + \dim(V') - 2 \dim(U' \cap V') \\
& = & (\dim(U') - \dim(U' \cap V')) \\
&& \quad + (\dim(V') - \dim(U' \cap V')) \\
& \le & ((\ell -  \kappa + \gamma) - (\ell - \kappa)) + (\ell - (\ell - \kappa)) \\
& = & \kappa + \gamma \; . 
\end{eqnarray*}

Note that $V' \in \code[m, \ell, k]$. The decoder $\decoder_{m, \ell, k}$ is able to correct 
any error pattern of size less than $\ell - k + 1$. Therefore, given that $\kappa+\gamma < \ell-k+1$, 
the decoder $\decoder_{m, \ell, k}$ will recover $V'$ from $U'$. Finally, $V$ is easily 
obtained as $ \{ \bldzero^\ell \} \oplus V'$. 
\qed

Next, we turn to estimate the time complexity of the decoder $\decoder_{\ell+m, \ell, k}$. 
We denote the decoding time of this decoder applied to $\code[\ell+m, \ell, k]$ as $T(\ell+m,\ell)$. 
Recall, that the time complexity of the algorithm $\decoder^\cK_{\ell+m, \ell, k}$ was shown 
in~\cite{SKK} to be $O(m (\ell-k))$ 
operations over $\ff_{q^m}$.
Then, the following recurrent relation holds:
\[
T (\ell+m, \ell) \le O \left( m (\ell-k) \right) + T (m, \ell)  \; . 
\]
The boundary condition is $T(\ell+m, \ell) = O(m (\ell-k))$ when $m < 2 \ell$. We obtain that 
\[
T (\ell+m, \ell) = O \left( \frac{m^2(\ell-k)}{\ell} \right) \le O \left( m^2 \right) \; . 
\]

\subsection{General Case}

Consider the case where $h_{\ell+m} \neq 0$ for some $m$, $\ell$. In this case, $|\cB| \le \ell$. Then, the algorithm in 
Figure~\ref{fig:alg} can be modified as follows. First, the decoder $\decoder^\cK_{\ell+m, \ell, k}$ is applied 
to the input $U \in W$. If that decoder fails, then $V \notin \cK$, and so 
one should look for $V \in \cB$ such that $d(U, V) < \ell-k+1$. 
This can be done by checking at most $O(\ell)$ ``candidates'' $V \in \cB$.
If one of them lies at the distance less than $\ell-k+1$ from $U$, then this $V$ is the codeword to be returned. 
If there is no such $V$, the failure `?' is returned.   

The resulting time complexity is a sum of the two following values:
\begin{itemize}
\item
Time complexity of the decoder $\decoder^\cK_{\ell+m, \ell, k}$.
\item
Time complexity of at most $\ell$ applications of the algorithm 
for computing distances between two vector subspaces.
\end{itemize}

Computing the distances between two vector subspaces can be 
done in a straight-forward manner by calculating the dimension 
of their intersection. It can also be done by an algorithm 
presented in~\cite[Sec. 3.4]{Silberstein}.

\section*{Acknowledgements}

The author would like to thank Danilo Silva and Emina Soljanin for helpful discussions,  
and Tuvi Etzion for pointing out the results in~\cite{Silberstein}. 
The author would also like to thank his colleagues in the Claude Shannon Institute: Carl Bracken, Eimear Byrne, Marcus Greferath, Russell Higgs, Nadya Markin, Gary McGuire and Alexey Zaytsev, --- 
for stimulating discussions and advices.

\end{document}